\documentclass[aps,prr,twocolumn,amssymb,superscriptaddress]{revtex4-2}
\usepackage{float}
\usepackage{graphicx}
\usepackage{epsfig}
\usepackage{epstopdf}
\usepackage{subfigure}
\usepackage{enumerate}
\usepackage{physics}
\usepackage{amssymb}
\usepackage{color}
\usepackage{rotating}
\newcommand{\RNum}[1]{\lowercase\expandafter{\romannumeral #1\relax}}
\begin{document}

\title{Fundamental trade-off between the speed of light and the Fano factor of photon current in three-level lambda systems}
\author{Davinder Singh}
\affiliation{Korea Institute for Advanced Study, Seoul 02455, Korea}
\author{Seogjoo J. Jang}
\affiliation{Korea Institute for Advanced Study, Seoul 02455, Korea}
\affiliation{Department of Chemistry and Biochemistry, Queens College, City University of New York, 65-30 Kissena Boulevard, Queens, New York 11367, USA  \& Chemistry and Physics PhD Programs, Graduate Center, City University of New York, 365 Fifth Avenue, New York, New York 10016, USA}
\author{Changbong Hyeon}
\thanks{hyeoncb@kias.re.kr}
\affiliation{Korea Institute for Advanced Study, Seoul 02455, Korea}

\begin{abstract}
Electromagnetically induced slow-light medium is a promising system for quantum memory devices, but controlling its noise level remains a major challenge to overcome. 
This work considers the simplest model for such medium, comprised of three-level $\Lambda$-systems interacting with bosonic bath, and provides a new fundamental trade-off relation in light-matter interaction between the group velocity of light and the Fano factor of photon current due to radiative transitions.  
Considering the steady state limits of a newly derived Lindblad-type equation, we find that the Fano factor of the photon current maximizes to 3 at the minimal group velocity of light, which holds true universally regardless of detailed values of parameters characterizing the medium. 
\end{abstract}
\maketitle

\section{Introduction} 
Quantitative characterization of fluctuations in driven quantum dynamical processes has fundamental implications for quantum thermodynamics \cite{esposito2009RMP,Millen_NJP18,Uzdin_PRX5,Taklner_RMP92,Mohammady_CP3,Miller_PRL123}, and is a central issue to address for the development of efficient quantum information \cite{Miller_PRL125,Preskill_Quantum2,DeLeon_Science372} and sensing devices \cite{Wasilewski_PRL104,McDonald_NC11,Yu_ACS_CS_7}.  
To this end, significant theoretical advances have been made in recent years, for example, by identifying new relations and bounds for stochastic/quantum fluctuations through quantum extensions \cite{esposito2009RMP,bruderer2014NJP,liu2019thermodynamic,hasegawa2021PRL,menczel2021thermodynamic,singh2021PRE,Carollo_PRL122}
 of thermodynamic uncertainty relations \cite{Seifert2012RPP,horowitz2019NaturePhys,Song2021JCP} and related quantum fluctuation theorems \cite{Deffner_PRL105,Mohammady_CP3,Miller_PRL123}.  
 As yet, utilizing many of these relations for actual experimental measurements/developments requires further theoretical analyses for establishing
concrete and experimentally testable relationships between physical observables.   This work provides such an analysis for a well 
known process that utilizes coherent driving of laser pulses to slow down light propagation \cite{hau1999Nature}, and 
clarifies an important trade-off relation in the process.

There have been considerable efforts to develop optical quantum memory devices employing laser control   \cite{budker1999nonlinear,ginsberg2007coherent,baba2008Natphoton,lvovsky2009NaturePhotonics,ma_optical_2017,goldzak2018PRL,li2020PRL} 
since Hau \emph{et al.} \cite{hau1999Nature} demonstrated extraordinary slowdown of the group velocity of light as slow as 17 m/s in an ultracold gas medium of sodium atoms. The electronic states of a sodium atom constitute a $\Lambda$-type three-level system, which comprises two nearly degenerate ground states and a common excited state.  
Applying a control pulse in resonance with the $\Lambda$-system can eliminate the linear absorption of a resonant probe pulse via destructive quantum interference, generating a \emph{dark state} where the atomic state is effectively trapped in the two ground states without excitation (see Appendix A for more precise description). 
Depending on the intensity of the control pulse relative to the probe pulse, two distinct mechanisms, coherent population trapping (CPT) \cite{gray1978OL,fu2005PRL} and electromagnetically induced transparency (EIT) \cite{electromagnetically_harris_1997}, make an otherwise absorbing medium effectively transparent and slow down the group velocity of the probe pulse propagating along the media of atomic vapor \cite{ma_optical_2017}. 
While conceptually clear, realization of an actual quantum memory device employing these phenomena 
has remained challenging due to a substantial level of noise \cite{hsu2006PRL,meyer2021PRA}. 
Although the major external sources of the noise have been identified and methods to suppress them have been developed over the years \cite{ma_optical_2017}, there still exist fluctuations \emph{inherent} in the radiative transitions generating photon currents. 
Elucidating the origin and size of these fluctuations 
under varying conditions could help understand the fundamental limit in achieving a given quantum memory device.

\begin{figure*}[ht!]
\includegraphics[width=0.9\linewidth]{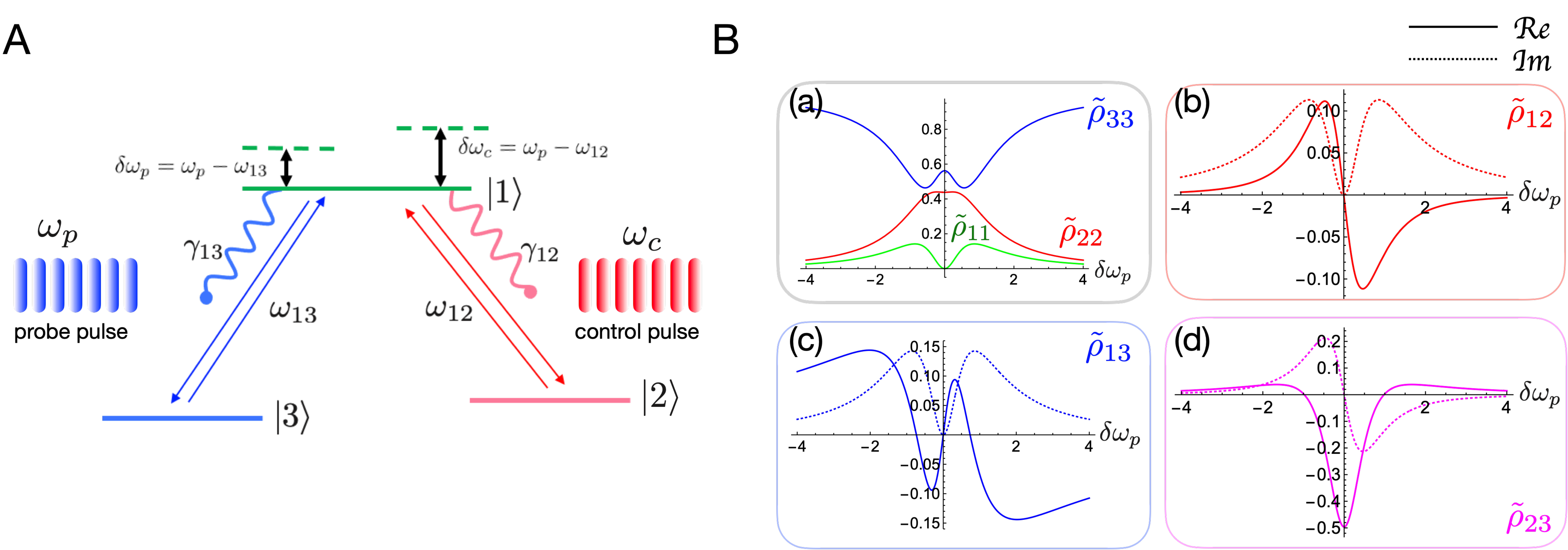}
\caption{Optical properties of $\Lambda$-system as a function of detuning frequency ($\delta\omega_p$). {\bf A.} Schematic of the system consisting of 3 electronic states, $\ket{1}$, $\ket{2}$ and $\ket{3}$, interacting with the probe and control pulses of frequencies $\omega_p$ and $\omega_c$. Here, $\omega_{12}(\equiv \omega_1-\omega_2)$ and $\omega_{13}(\equiv \omega_1-\omega_3)$ are the resonant frequencies. 
Further, $\delta\omega_c = \omega_c - \omega_{12}$ and $\delta\omega_p = \omega_p - \omega_{13}$ denote the detuning frequencies. 
The condition $\delta\omega_p=\delta\omega_c=0$ corresponds to the two-photon resonance.  
{\bf B.} Populations in $\ket{1}$, $\ket{2}$, and $\ket{3}$ are shown in the panel (a). 
Real and imaginary parts of the coherences 
$\tilde{\rho}_{12}$, $\tilde{\rho}_{13}$, and $\tilde{\rho}_{23}$ are depicted in (b), (c), and (d) as a function of $\delta\omega_p$ with the solid and dotted lines, respectively. 
Here, we have used $\gamma\equiv\gamma_{12}/\gamma_{13}= 0.9$, $\delta\omega_c = 0$, $\bar{n}_{ij} = 0$, $\Omega_c= 0.56$, and $\Omega_p = 0.50$. 
All the frequencies are scaled with $\gamma_{13}(\approx 0.62\times 10^8$ s$^{-1}$).  
}
\label{fig:coherences}
\end{figure*}

The main objective of this work is to offer a quantitative understanding of how the relative fluctuations of photon current associated with radiative transitions in a coherently controlled ensemble of $\Lambda$-systems change as the group velocity of light is reduced. 
In a recent work on a field-driven two-level system (TLS) weakly interacting with bosonic environment \cite{singh2021PRE}, we have shown that the Fano factor (or relative fluctuations) of photon current associated with radiative transitions is determined by the competition between the real and imaginary parts of the steady state coherence formed between the excited and ground states, such that the imaginary part of the coherence reduces the fluctuations, whereas the real part contributes to enhancing them \cite{singh2021PRE}. 
Employing a similar formalism for the $\Lambda$-system and through careful theoretical analyses of 
a Lindblad-type equation while treating light-matter interaction at semi-classical level, we discover a fundamental trade-off relation between the speed of light and the Fano factor of photon current.

\section{Theoretical model}
A three-level $\Lambda$-system comprised of the electronic states $\ket{1}$, $\ket{2}$, and $\ket{3}$ is coupled to a thermally-equilibrated bosonic bath at temperature $T$. 
The system is illuminated with control ($\alpha=c$) and probe ($\alpha=p$) laser pulses, 
$\vec{E}_\alpha({\bf r},t)=\hat{\epsilon}_\alpha\zeta_\alpha(e^{-i({\bf k}_\alpha\cdot{\bf r}-\omega_\alpha t)}+e^{i({\bf k}_\alpha\cdot{\bf r}-\omega_\alpha t)})\simeq \hat{\epsilon}_\alpha\zeta_\alpha(e^{i\omega_\alpha t}+e^{-i\omega_\alpha t})$,  
each with the amplitude $\zeta_\alpha$, wave vector ${\bf k}_\alpha$, and the angular frequency $\omega_\alpha$. 
The two polarization vectors, $\hat{\epsilon}_c$ and $\hat{\epsilon}_p$ are orthogonal to each other ($\hat{\epsilon}_c\cdot\hat{\epsilon}_p=0$), and the dipole approximation (${\bf k}_{\alpha}\cdot{\bf r}\ll1$) \cite{scully_zubairy_1997} is taken at the second equality of $\vec{E}_\alpha({\bf r},t)$ since the atomic length scale is much smaller than the wavelength of laser pulses. 
In addition, we simplify the situation here by focusing on the linear response regime \cite{ma_optical_2017,Fleischhauer_RMP77} with respect to the probe field and on the dilute sample limit where collective excitation or multiple atom-light scattering does not make significant contribution. 
The full Hamiltonian representing this model is provided in Appendix B. 

The atoms in $\ket{2}$ and $\ket{3}$ states are excited to a common excited state $\ket{1}$ through interactions of  transition dipole opertors, $\vec{d}_2$ (between $\ket{1}$ and $\ket{2}$) and $\vec{d}_3$ (between $\ket{1}$ and $\ket{3}$), with the incident pulses (see Fig.~\ref{fig:coherences}A).  This is  represented by an interaction Hamiltonian $H_{\rm int}=-\vec{d}_2\cdot \vec{E}_c-\vec{d}_3\cdot\vec{E}_p$, for which 
two Rabi frequencies $\Omega_c$ and $\Omega_p$ characterizing the respective interaction strengths can be defined (see Appendix B for details). 
The state $\ket{1}$ can either decay into $\ket{2}$ with a rate $\gamma_{12}$ or into $\ket{3}$ with $\gamma_{13}$.  The transition between $\ket{2}$ and $\ket{3}$ is effectively spin-disallowed with $\gamma_{23}\ll \gamma_{12}$, $\gamma_{13}$. 
Employing the standard assumptions of the weak system-bath coupling, Born-Markov, and the rotating wave approximations (RWA), we find that the dynamics of the $\Lambda$-system can be described by the following Lindblad-type equation for the reduced density matrix $\rho(t)$ (see Appendix B), 
\begin{align}
\partial_t\rho(t)=-(i/\hbar)[H_S+H_{\rm int},\rho(t)]+\mathcal{D}(\rho(t)),
\label{eqn:evolution}
\end{align}
where $H_S=\hbar(\omega_1\ket{1}\bra{1}+\omega_2\ket{2}\bra{2}+\omega_3\ket{3}\bra{3})$ with $\hbar\omega_i$ denoting the energy level of the $i$-th state, and $\mathcal{D}(\rho(t))$ is a Lindblad-type dissipator. 
Note that there are multiple ways to formulate the phenomenon of slow light. For example, one can study the light-matter interaction by explicitly quantizing the electric field as well as the atomic state, but either by ignoring the effect of bath \cite{arkhipov2022PRL} or by treating the effect of bath only phenomenologically \cite{fleischhauer2000dark}. 
Our formulation in this study rests on a Lindblad-type equation that explicitly takes into account the effect of fast relaxing background photon bath on the system, but treats the interaction with primary control and probe pulses at semi-classical level. 

Equation~(\ref{eqn:evolution}) can be transformed to  
$\partial_t\tilde{\varrho}(t)=\mathcal{L}\tilde{\varrho}(t)$
where $\tilde{\varrho}\equiv(\tilde{\rho}_{11},\tilde{\rho}_{12},\tilde{\rho}_{13},\tilde{\rho}_{21},\tilde{\rho}_{22},\tilde{\rho}_{23},\tilde{\rho}_{31},\tilde{\rho}_{32},\tilde{\rho}_{33})^T$ is vector representation of $\rho(t)$ in the rotating wave frame (see Appendix C), 
and $\mathcal{L}$ represents the Liouvillian super-operator expressed as $9\times 9$ matrix in the Fock-Liouville space \cite{manzano2020AIP}. 
The steady-state value of each element $\tilde{\rho}^{ss}_{ij}$ is calculated from $\mathcal{L}\tilde{\varrho}^{ss}=0$ (see Eq.~(\ref{coh_real})).  
Fig.~\ref{fig:coherences} shows the population in each state ($\tilde{\rho}^{ss}_{ii}$, which satisfies $\sum_{i=1,2,3}\tilde{\rho}_{ii}^{ss}=1$) and coherences between the states $\ket{i}$ and $\ket{j}$ ($\tilde{\rho}^{ss}_{ij}=\rho^R_{ij}+i\rho^I_{ij}$, $i\neq j$, with $\rho^R_{ij}\equiv\Re{\tilde{\rho}_{ij}^{ss}}$ and $\rho^I_{ij}\equiv\Im{\tilde{\rho}_{ij}^{ss}}$) as a function of the detuning frequency of the probe pulse ($\delta\omega_p$).  

The condition of two-photon resonance ($\delta\omega_p=\delta\omega_c=0$) and $\Omega_c\approx \Omega_p$ engender a special atomic state termed a \emph{dark state}: 
the atom is locked in the states $\ket{2}$ and $\ket{3}$, without populating the excited state $\ket{1}$, i.e., $\tilde{\rho}_{22}$, $\tilde{\rho}_{33}\neq 0$ but $\tilde{\rho}_{11}=0$ (panel (a) of Fig.~\ref{fig:coherences}B). 
In addition, except for the real part of the coherence between $\ket{2}$ and $\ket{3}$ ($\rho_{23}^R\neq 0$), all the coherence terms vanish, such that $\rho_{12}^R = \rho_{12}^I = \rho_{13}^R = \rho_{13}^R = \rho_{23}^I = 0$. 
This situation corresponds to the CPT, where the effects of control and probe pulses are cancelled off via destructive interference, and 
the atomic state is delocalized between $\ket{2}$ and $\ket{3}$, forming 
\emph{a dark state}. 
It is also noteworthy that in the dark state, both the photon current between the atomic states and its variance vanish; yet their ratio corresponding to the Fano factor remains finite, which constitutes the major result of our work. 
Since there is neither dispersion ($\rho_{13}^R =0$) nor absorption of light ($\rho_{13}^I=0$), the atomic medium looks effectively transparent to the probe pulse (see Appendix A for more complete description of the dark state, CPT and EIT).

\section{Photon current, fluctuations, and Fano factor}
Laser pulse applied to the system for a time interval sufficiently longer than the decay time ($\tau\equiv \gamma_{13}t\gg 1$) establishes steady-state current of photon absorption and emission. 
With the net number of radiative transitions in the $\Lambda$-system denoted as $n(\tau)$, where $n(\tau)>0$ is for emissions and $n(\tau)<0$ is for absorptions,
the average photon current at steady state ($J_{\rm ph}$), its variance ($D_{\rm ph}$), and the corresponding Fano factor ($\mathcal{F}$) are defined as follows. 
\begin{align}
&J_{\rm ph} \equiv \lim_{\tau\gg 1}\frac{\langle n(\tau) \rangle}{\tau},\nonumber\\
&D_{\rm ph} \equiv \lim_{\tau\gg 1}\frac{{\rm var}[n(\tau)]}{\tau}, \nonumber\\
\mathcal{F} &=\frac{D_{\rm ph}}{J_{\rm ph}}=\lim_{\tau\gg 1} \frac{{\rm var}[n(\tau)]}{\langle n(\tau)\rangle},  
\label{fano}
\end{align}
where ${\rm var}[n(\tau)]\equiv \langle n(\tau)^2\rangle-\langle n(\tau)\rangle^2$. 
Detailed expressions of these for the $\Lambda$-system can be obtained by employing the method of cumulant generating function \cite{flindt2010PRB,bruderer2014NJP} (see Appendix D).  

\begin{figure}[t!]
\includegraphics[width=1\linewidth]{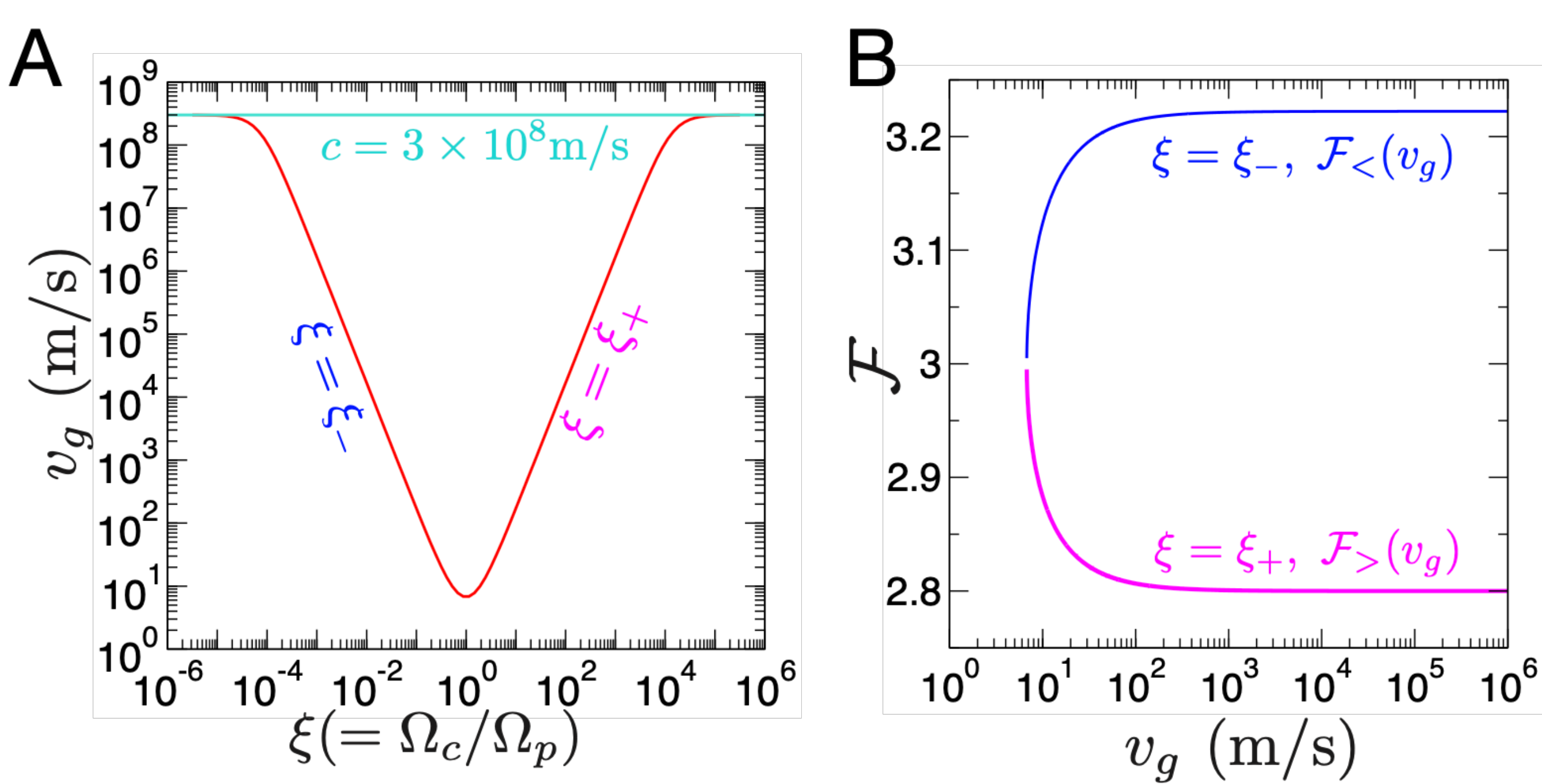}
\caption{Group velocity ($v_g$) and Fano factor ($\mathcal{F}$). 
{\bf A.} $v_g=v_g(\xi)$ in red, and vacuum speed of light $c$ in blue.  
{\bf B.} $\mathcal{F}$ versus $v_g$ calculated by varying $\xi(=\Omega_c/\Omega_p)$ at two-photon resonance ($\delta\omega_p = \delta\omega_c = 0$).  
Depending on whether $\xi<1$ or $\xi>1$, $\mathcal{F}$ changes differently with $v_g$. 
For the calculation, the parameters were taken from Hau \emph{et al.} \cite{hau1999Nature} that experimented on $^{23}$Na atom: {\color{black}$\bar{n}_{ij} \approx 0$} ($\mathcal{A}\gg 1$),  
$\gamma(\equiv \gamma_{12}/\gamma_{13})=0.9$, and $\mathcal{N}=2\pi N_d(\omega_p/\Omega_p)\approx 1.78\times 10^8$, which is estimated from 
$N_d=N|\vec{d}_{13}|^2/(\hbar\Omega_p\gamma_{13})=0.11$ 
with $N\approx 8 \times 10^{13}$ cm$^{-3}$, $|\vec{d}_{13}|\approx 1.4\times 10^{-29}$ C$\cdot$m $\approx4.2\times 10^{-18}$ statC$\cdot$cm, $\Omega_p=0.2$ \cite{steck2003sodium}, $\gamma_{13}\approx0.62 \times 10^8$ s$^{-1}=(16.23 \text{ ns})^{-1}$, and $\omega_p=(2\pi c/\lambda_p)/\gamma_{13}\approx 2 \pi\times 8.21\times 10^6$ with $\lambda_p\approx 589$ nm.}
\label{GroupVel_fano}
\end{figure}

When the two energy gaps are identical ($\omega_{12} = \omega_{13} = \omega_0 $), 
the mean number of background thermal photons at this frequency is given by $\bar{n}_{12} = \bar{n}_{13} = \bar{n}=(e^{\beta\hbar\omega_0}-1)^{-1}$. 
Then, $\mathcal{F}$ simplifies to   
(see Appendices D and E)
\begin{align}
\mathcal{F} &= \coth{\left( \frac{\mathcal{A}}{2} \right)}\left[1+\mathcal{R}-\mathcal{I}+q(\cdot)\right], 
\label{fano1}
\end{align}
where $\mathcal{A} = \beta\hbar\omega_0$, $\mathcal{R}\equiv 2\sum_{i\neq j}\left(\rho_{ij}^R\right)^2$, $\mathcal{I}\equiv 6\sum_{i\neq j}\left(\rho_{ij}^I\right)^2$ with $i,j\in\{1,2,3\}$, and $q(\cdot)=q(\Omega_c,\Omega_p,\gamma,\mathcal{A},\delta\omega_c,\delta\omega_p)$. 
Similarly to the Fano factor for the field-driven TLS \cite{singh2021PRE}, 
$\mathcal{F}$ of the $\Lambda$-system is determined by the competition between the real ($\mathcal{R}$) and imaginary ($\mathcal{I}$) parts of steady-state coherence; however, there is an additional factor $q(\cdot)$ in the expression (Eq.~(\ref{fano1})), which is absent in the TLS but could be significant in determining the magnitude of $\mathcal{F}$ for the $\Lambda$-system.  
The full expression of $q(\cdot)$ is rather complicated, but at the two-photon resonance it is greatly simplified to 
\begin{align}
q(\cdot)=\frac{2(\gamma\xi^6+2\gamma\xi^4+2\xi^2+1)}{(\xi^2+1)(\xi^2+\gamma)^2}, 
\label{eqn:qval}
\end{align}
where $\xi(\equiv\Omega_c/\Omega_p)$ is the experimentally controllable variable, and $\gamma\equiv \gamma_{12}/\gamma_{13}$ 
(see Eqs.~(\ref{eqn:q_org}) and (\ref{eqn:q_0})). Note that  the result of TLS, i.e., $q(\cdot)=0$ is recovered under the limiting condition of $\gamma\gg 1$.

{\it Group velocity of probe field and Fano factor. }
Since the group velocity of light is defined as $v_g=\left[dk(\omega)/d\omega\right]^{-1}$, where $k(\omega)=\omega \eta(\omega)/c$ with $\eta(\omega)$ denoting the real part of the refractive index and $c$ speed of light in vacuum,
a change in the refractive index gives rise to a change in the group velocity of probe field across the medium as follows (see Appendix F)
\begin{align}
v_g 
&=c\left(\eta(\omega)+\omega\frac{d\eta (\omega)}{d\omega}\right)^{-1}\nonumber\\
&= \frac{c}{1 + 2\pi N_d\rho_{13}^R + 2\pi\omega_pN_d(\partial\rho_{13}^R/\partial\omega_p)}. 
\label{group_vel}
\end{align}
where $N_d\equiv N|\vec{d}_{13}|/\zeta_p(=N|\vec{d}_{13}|^2/\hbar\Omega_p\gamma_{13})$ with $N$ being the  density of atoms comprising the medium of atomic vapor. 

\begin{figure*}[ht]
\includegraphics[width=0.75\linewidth]{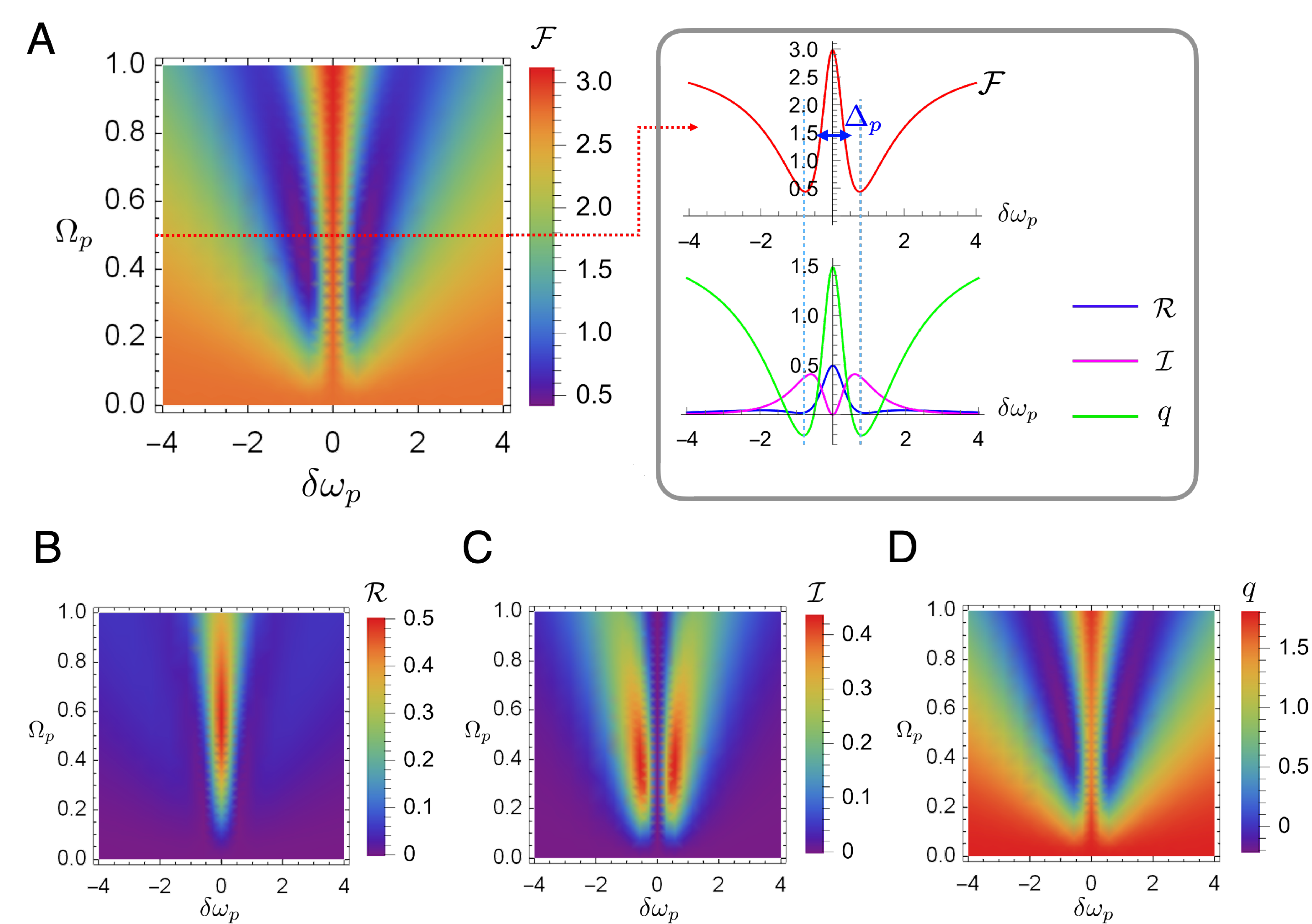}
\caption{Effect of detuning on the Fano factor of radiative transitions. 
{\bf A.} Diagram of $\mathcal{F}(\delta\omega_p,\Omega_p)$ calculated for $\delta\omega_c=0$ with $\Omega_c = 0.56$, $\gamma = 0.90$, $\mathcal{A} = 47$.
(Inset) $\mathcal{F}$, $\mathcal{R}$, $\mathcal{I}$, and $q$ as a function of $\delta\omega_p$ for $\Omega_p=0.5$. The blue vertical dashed line indicates the value of $\delta\omega_p(\approx 0.8)$ that gives rise to the minimal $\mathcal{F}$. 
The range of transparency window ($\Delta_p$) is indicated by the arrow.  
{\bf B.} Real ($\mathcal{R}$) and  
{\bf C.} imaginary parts of coherence ($\mathcal{I}$)  and 
{\bf D.} the factor $q$ as a function of probe detuning $\delta\omega_p$ and  driving frequency $\Omega_p$. 
}
\label{F_calc}
\end{figure*}

The condition of two-photon resonance ($\delta\omega_p = \delta\omega_c = 0$) simplifies Eq.~(\ref{group_vel}) with $(\rho_{13}^R)_{\delta\omega_p=0} = 0$ (Fig.~\ref{fig:coherences}B, Fig.~\ref{F_calc}A inset, and see Eq.~(\ref{coh_real})). 
Hence, $v_g$ is greatly reduced by increasing the derivative term, $(\partial\rho_{13}^R/\partial\omega_p)_{\delta\omega_p=0}$, namely, by increasing the variation of refractive index (or coherence) involving the states $\ket{1}$ and $\ket{3}$ with respect to the probe pulse frequency, $\omega_p$ \cite{hau1999Nature}. 
In fact, it is straightforward to show $(\partial\rho_{13}^R/\partial\omega_p)_{\delta\omega_p=0}=\Omega_p^{-1}(\rho^R_{23})^2_{\delta\omega_p=0}$ (Eq.~(\ref{coh_real})). Thus, $v_g$ in Eq.~(\ref{group_vel}) is determined by the strength of Raman coherence, i.e., the magnitude of the real part of coherence between the two ground states $\ket{2}$ and $\ket{3}$ at two-photon resonance ($\delta\omega_c=\delta\omega_p=0$) as follows. 
\begin{align}
v_g=\frac{c}{1+\mathcal{N} (\rho^R_{23})^2_{\delta\omega_p=0}},  
\label{eqn:vg}
\end{align}
where $\mathcal{N}\equiv 2\pi N_d\omega_p/\Omega_p$ is a factor determined by the density of atoms comprising the medium, the magnitude of the transition dipole moment $|\vec{d}_{13}|$, the resonant and Rabi frequencies, $\omega_p$ and $\Omega_p$.

An important relation between $v_g$ and $\mathcal{F}$ for $\Lambda$-systems can be identified through $\xi$ (see Fig.~\ref{GroupVel_fano}A for $v_g=v_g(\xi)$). 
Fig.~\ref{GroupVel_fano}B shows a curve of $\mathcal{F}$ versus $v_g$ parameterized with $\xi$ at $\delta\omega_p=\delta\omega_c=0$ for $\gamma=0.9$, 
clarifying a trade-off relation between $\mathcal{F}$ and $v_g$ for experimentally relevant range of variable, $\xi>1$. 
It is noteworthy that the Fano factor of photon transitions sharply increase to $\mathcal{F}\simeq 3$ when $v_g$ approaches its minimal value $v_g\simeq 7$ m/s 
(Fig.~\ref{GroupVel_fano}B, magenta line), which is even smaller than the one experimentally reported \cite{hau1999Nature}.

For $\mathcal{A}\gg 1$ (or $\bar{n} \sim 0$) with $\delta\omega_c = \delta\omega_p=0$, 
the expressions of coherence terms (Eq.~(\ref{coh_real})) are greatly simplified, enabling us to further clarify a relation between $v_g$ and $\mathcal{F}$. 
With $\left(\rho_{23}^R\right)^2_{\delta\omega_p=0}=\xi^2/(\xi^2+1)^2$,  
$\rho_{23}^I=\rho_{12}^R=\rho_{12}^I=\rho_{13}^R=\rho_{13}^I=0$ (Eq.~(\ref{coh_real})) and 
the expression of $q(\cdot)$ given in Eq.~(\ref{eqn:qval}), 
the group velocity and the Fano factor read  
\begin{align}
v_g=\frac{c}{1+\dfrac{\mathcal{N}}{(\xi+1/\xi)^2}}
\label{eqn:velocity}
\end{align} 
and 
\begin{align}
\mathcal{F}\simeq 1+\frac{2(1+\gamma \xi^2)}{(\gamma+\xi^2)}. 
\label{eqn:Fano}
\end{align}
From Eq.~(\ref{eqn:velocity}), it is clear that 
$v_g$ minimizes to $v_g^\text{min}=c/(1+\mathcal{N}/4)$ for $\xi=1$, and saturates to $v_g=c$ for $\xi \gg \sqrt{\mathcal{N}}$ or $\xi\ll 1/\sqrt{\mathcal{N}}$ (see Fig.~\ref{GroupVel_fano}A). 
Next, the term $\xi$ in Eq.~(\ref{eqn:velocity}) 
can be solved in terms of $v_g$, yielding two expressions, $\xi=\xi_{\pm}=\frac{1}{2}[\sqrt{\mathcal{N}/(c/v_g-1)}\pm\sqrt{\mathcal{N}/(c/v_g-1)-4}]\gtrless 1$. Insertion of $\xi=\xi_\pm$ to Eq.~(\ref{eqn:Fano}) 
yields $\mathcal{F}=\mathcal{F}_>(v_g)$ for $\xi=\xi_+(>1)$ (magenta line in Fig.~\ref{GroupVel_fano}B),  and $\mathcal{F}=\mathcal{F}_<(v_g)$  for $\xi=\xi_-(<1)$ (blue line in Fig.~\ref{GroupVel_fano}B). 
We note that only the condition of $\xi>1$ is of practical relevance to the slow-light experiment because the current fluctuations are smaller and more controllable with $\mathcal{F}_>(v_g)\leq 3$.  
At $\xi=1$ or equivalently at $v_g=v_g^\text{min}$, one always obtains $\mathcal{F}=3$.  The universality of this value is a key outcome of our analyses.

For more general case with $\delta\omega_p\neq 0$ and $\delta\omega_c=0$, 
the expression of $\mathcal{F}$ is complicated; yet, $\mathcal{F}$ is still an even function of $\delta\omega_p$ (Eq.~(\ref{coh_real})). 
Confining ourselves to the condition $\xi>1$, we resort to numerics to calculate $\mathcal{F}(\delta\omega_p,\Omega_p)$ (Fig.~\ref{F_calc}), finding that $\mathcal{F}$ is maximized over the transparency window $\Delta_p$, given by 
$\Delta_p\sim \left[\partial\rho_{13}^R/\partial\delta\omega_p\Big|_{\delta\omega_p=0}\right]^{-1}=\Omega_p(\xi^2+1)^2/\xi^2$. 
Note that $\Delta_p$ is narrow for the case of CPT ($\xi\approx 1$) but is wide for EIT ($\xi\gg 1$). 
Over the narrow transparency window $\Delta_p$, 
the coherence between atomic states $\ket{1}$ and $\ket{3}$ vanish ($\rho_{13}^R$, $\rho_{13}^I\approx 0$) (Fig.~\ref{fig:coherences}B-(c)), 
and $\mathcal{R}$ and $q$ display maximal contribution at two-photon resonance (Fig.~\ref{F_calc}A inset, B and D), 
whereas $\mathcal{I}\approx 0$, {\it i.e.}, the absorption is negligible (Fig.~\ref{F_calc}A inset and C). 

It is worth noting that the Fano factor of radiative transitions is maximally reduced under a detuning condition $\delta\omega_p\neq 0$ 
where $\mathcal{I}$ is maximized, $\mathcal{R}\approx 0$, and $q(\cdot)<0$, resulting in $\mathcal{F} < 1$ (Fig.~\ref{F_calc}A inset, B, and D); 
however, such a condition is attained when the value of $\delta\omega_p$ is beyond the transparency window, 
which does not correspond to the regime where absorption-free slow light can be generated. 
Rather, under such condition, the absorption doublet arises from the transitions from $\ket{0}$ to two eigenstates $\ket{\pm}$ comprised of the three electronic states $\ket{1}$, $\ket{2}$, and $\ket{3}$ \cite{scully_zubairy_1997} (see Fig.~\ref{EIT}B and Eq.~(\ref{eqn:eigenstates})).

\section{Concluding Remarks}
This work, which considers a model of a coherently controlled $\Lambda$-type three-level system interacting with thermalized background photons, has established a fundamental trade-off  relation between the group velocity of light and the Fano factor of photon current of the radiative transition in electromagnetically induced slow light medium. 
In particular, the Fano factor of the net number of radiative transitions $n(\tau)$, which dictates the relative 
fluctuations of the laser power (see Appendix H, $\langle (\delta n(\tau))^2\rangle/\langle n(\tau)\rangle \propto\langle (\delta I)^2\rangle/\langle I\rangle$), is maximized to $\mathcal{F}=3\coth{(\mathcal{A}/2)}$ at the slowest group velocity, $v_g\approx (4/\mathcal{N})c$. 
This indicates that slow light is attained at the expense of relative fluctuations of the irreversible photon current. 
This trade-off, which may be inevitable in the basic setup of CPT or EIT-based optical quantum memory device, is physically sensible in that as the light slows down, overall fluctuations in the photon current is enhanced over the prolonged travel time of the photon inside the medium. 
At two-photon resonance, 
the real part of coherence between the two ground states ($\rho_{23}^R$), which engenders slow light (Eq.~(\ref{eqn:vg})) and increases the Fano factor of signal (Eq.~(\ref{fano1})), is maximized at the regime corresponding to CPT, where the Rabi frequencies of control and probe pulses are identical ($\xi=\Omega_c/\Omega_p=1$).

Our results can also be applied to the medium consisting of $^{133}$Cs atoms, one of two major systems being used currently for EIT quantum memory scheme \cite{ma_optical_2017}, whose $D1$ line constitutes the three-level $\Lambda$-system. 
For Cs atoms, the frequency gap between the two ground states $6^2S_{1/2}(\ket{F=3})$ and $6^2S_{1/2}(\ket{F=4})$, where $F$ stands for the total angular momentum quantum number,  
is $\sim$ 9.2 GHz. 
The condition of $\rho_{23}^R\neq 0$ and $\rho_{23}^I=0$ signifies a Raman coherence between $\ket{F=3}$ and $\ket{F=4}$ effectively with no absorption. 
The slowest group velocity achievable for the case of CPT regime ($\xi\approx 1$) of $^{133}$Cs vapor \cite{hockel2010PRL} 
is $v_g\approx 38$ m/s with $\mathcal{N}=2\pi N_d(\omega_p/\Omega_p)\approx 3.2\times 10^7$, which is estimated from $\Omega_p=0.5$, $\omega_p=(2\pi c/\lambda_p)/\gamma_{13}\approx 2.1\times 10^7$ with $\lambda_p\approx 894$ nm \cite{hockel2010PRL} and $\gamma_{13}\approx 10^8$ s$^{-1}$, and 
$N_d=N|\vec{d}_{13}|^2/(\hbar\Omega_p\gamma_{13})=0.12$ 
with $N\approx 10^{12}$ cm$^{-3}$ and $|\vec{d}_{13}|=2.7\times 10^{-29}$ C$\cdot$m $=8.09\times 10^{-18}$ statC$\cdot$cm \cite{steck2003cesium}.   
It is important to note that our estimate for the slowest group velocity of light in the atomic vapor of cesium is amenable for an experimental verification. 

Our main result concerning the size of the relative fluctuations (Fano factor) of photon current (or noise level) due to radiative transitions of three-level $\Lambda$-system at the slowest group velocity is universal ($\mathcal{F}=3$) regardless of the atomic type, which warrants experimental confirmation.  
Our theory is formulated for the storage process, but not explicit in addressing the fluctuations of signal upon retrieval. Yet, it is still known from direct experimental measurements that the photon number statistics are preserved during the storage and retrieval processes \cite{cho2010PRA}. Thus, the noise level at the storage process discussed in this study is expected to carry over to the retrieved signal as well.
The formalism of this work can be extended to other 
types of systems, for example, with  $V$ and ladder structures \cite{ma_optical_2017,tscherbul2014PRL,Koyu_PRR3,zanner2022NatPhys} 
and also to Bose-Einstein condensates that can serve as media where the light can stop completely \cite{ginsberg2007coherent}. 
However, in actual experimental situations, some effects that are not accounted for by our model may have nontrivial effects.  
For example, there could be 
cases where control or probe field 
interacts with another nearby energy level \cite{hau1999Nature}, 
resulting in additional decoherence mechanism. 
Within our model, such an effect could in principle be incorporated 
by modifying the $\rho_{23}$-involving term in Eq.~(\ref{eqn:vg}), which would lead to an observed group velocity deviating from the fundamental limit predicted by Eq.~(\ref{eqn:vg}). 
More challenging cases are when the effects of collective emission \cite{Svidzinsky_PRA81} or multiple scattering effects \cite{Oliveira_PRA104} are significant, for which formulation that goes beyond our model becomes necessary. 
Another important theoretical challenge is treating probe and control fields fully quantum mechanically. 
How the trade-off relation is altered for the different systems and by additional effects due to non-Markovian or strongly coupled environments \cite{tanimura2020JCP,ikeda2020ScienceAdvances,rivas2020PRL} remains an important theoretical issue that requires further investigation.  

\begin{acknowledgements}
We thank Prof. Hyukjoon Kwon for careful reading of the manuscript and helpful discussions. 
This work was supported by the KIAS Individual Grants (CG077602 to DS, and CG035003 to CH) from the Korea Institute for Advanced Study, and by the US National Science Foundation (CHE-1900170 to SJJ).
We thank the Center for Advanced Computation in KIAS for providing computing resources.
\end{acknowledgements}



\section*{Appendix}

\setcounter{figure}{0}
\renewcommand{\thefigure}{A\arabic{figure}}
\setcounter{equation}{0}
\renewcommand{\theequation}{A\arabic{equation}}

\subsection{Coherent population trapping (CPT) and electromagnetically induced transparency (EIT)}

\begin{figure}[b]
\includegraphics[width=1.0\columnwidth]{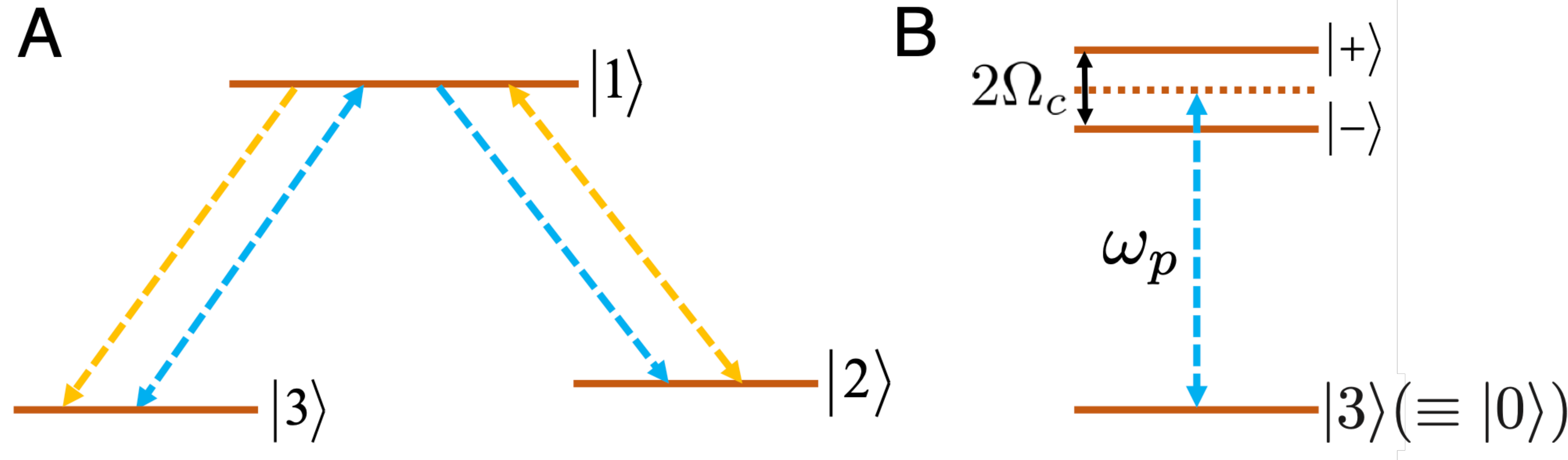}
\caption{{\bf A.} Bare state basis to show the paths involved in the destructive interference for 
$\Omega_c/\Omega_p\approx 1$.
and {\bf B.} the corresponding dressed state picture for the weak probe field ($\Omega_c/\Omega_p\gg 1$).
}
\label{EIT}
\end{figure}

{\bf CPT.} The absorption and dispersion profiles of probe pulse as a function of  detuning ($\delta\omega_p$) are calculated in Fig.~\ref{fig:coherences}B in the main text. 
At the two-photon resonance ($\delta\omega_p=\delta\omega_c = 0$), 
both the coherences between the states $\ket{1}$ and $\ket{3}$ and between the states $\ket{1}$ and $\ket{2}$ vanish  
($\rho_{13}^R=\rho_{13}^I=0$ and $\rho_{12}^R=\rho_{12}^I=0$ in Fig.~\ref{fig:coherences}B), 
which implies that the medium is effectively transparent to the probe and control pulses. 
The two light pulses interacting with the matter vanish via the destructive interference between two pathways between $\ket{3} \rightleftharpoons \ket{1} \rightarrow \ket{2}$ and $\ket{2} \rightleftharpoons \ket{1} \rightarrow \ket{3}$ (Fig.~\ref{EIT}A).

\begin{figure}[ht]
\includegraphics[width=0.8\columnwidth]{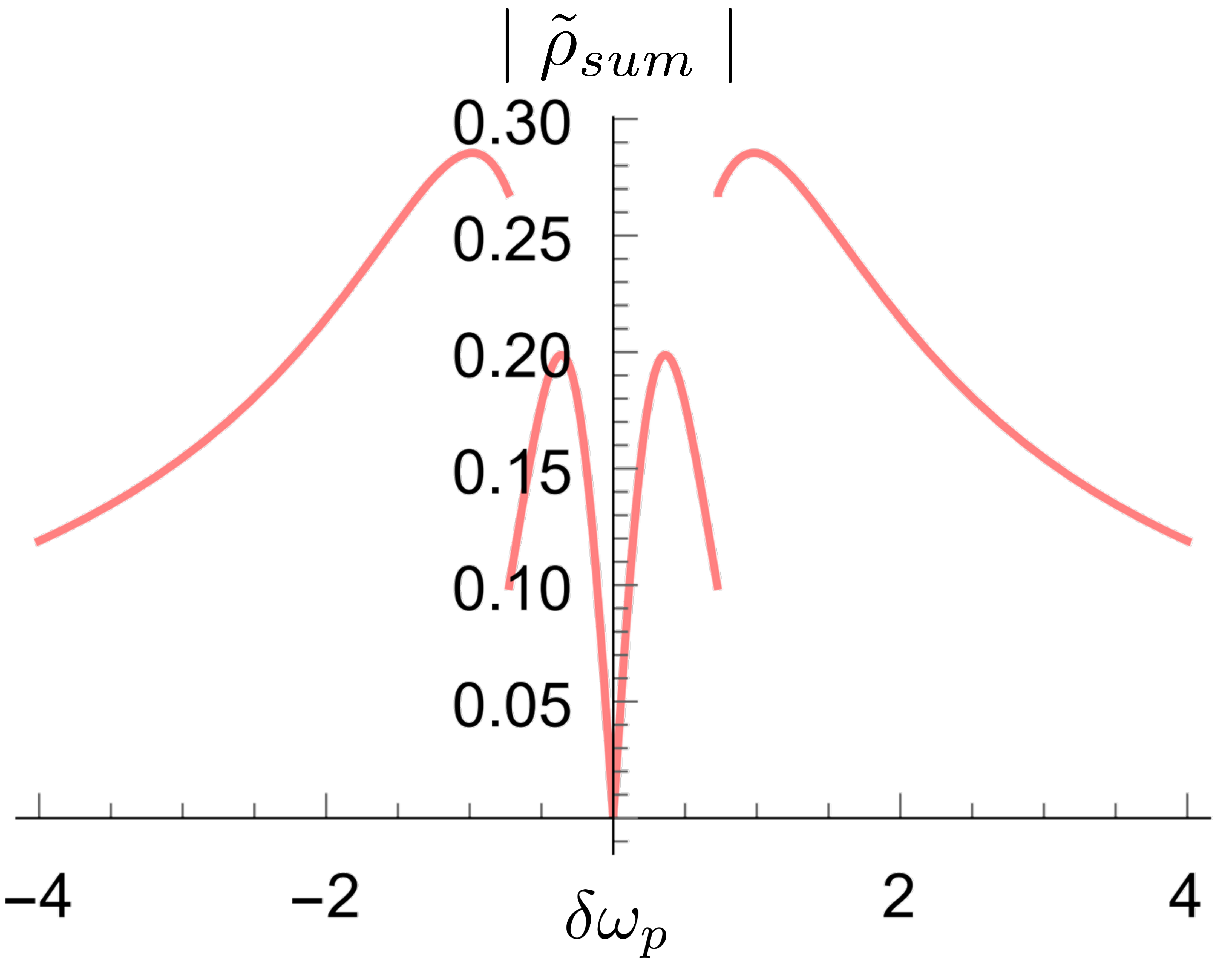}
\caption{ Plot of $|\tilde{\rho}_{sum}| = |\tilde{\rho}_{12} + \tilde{\rho}_{13}|$ with varying $\delta\omega_p$ with fixed $\delta\omega_c = 0$ for $\Omega_c= 0.56$, $\Omega_p = 0.50$, $\gamma = 0.9$, $\bar{n}_{ij} = 0$.
}
\label{Phase_relation}
\end{figure}

To show the destructive quantum interference more explicitly, we consider an addition of two pulses with quantum coherence,
\begin{align}
\tilde{\rho}_{\rm sum} = \tilde{\rho}_{12} + \tilde{\rho}_{13}. 
\end{align}
Note that $\tilde{\rho}_{ij} = | \tilde{\rho}_{ij} | \text{exp}(i\theta_{ij})$ with $| \tilde{\rho}_{ij} |^2 = \left(\rho_{ij}^R\right)^2 + \left(\rho_{ij}^I\right)^2 $ and $\tan{\theta_{ij}} = \left(\rho_{ij}^I/\rho_{ij}^R \right)$.
Numerical calculation using the results in Eq.~(\ref{coh_real}) 
gives rise to Fig.~\ref{Phase_relation}, indicating that the amplitude of $\tilde{\rho}_{\rm sum}$ vanishes at two-photon resonance ($\delta\omega_p = \delta\omega_c = 0$). 
Thus, the excitation transfer to the state $\ket{1}$, and hence the photon current, is negligible, and almost all the atomic population is trapped in the  states $\ket{2}$ and $\ket{3}$ (Fig.~\ref{fig:coherences}A in the main text). 
The ``coherent population trapping'' (CPT) refers to such a trapping of atomic population in the two ground states via a coherent superposition of the quantum states.

The destructive interference and hence population trapping in states $\ket{2}$ and $\ket{3}$ results in strong coupling between these states, which is reflected in the high value of $\rho_{23}^R$ (see Fig.~\ref{fig:coherences}B in the main text).

More complete physical interpretation of CPT can be given in terms of the basis representing the dressed (or eigen) states. 
Under the following unitary transformation, which is equivalent to describing the system in the rotating frame,  
\begin{align}
\ket{\psi} =\mathcal{U} \ket{\phi},
\end{align}
where $\mathcal{U}=e^{-i\omega_p t \ket{1}\bra{1} - i\left( \omega_p - \omega_c \right)t \ket{2}\bra{2}}$, 
the Schr{\"o}dinger equation $\partial_t\ket{\psi} = -iH/\hbar \ket{\psi}$ is written as $\partial_t\ket{\phi} = -iH_{\rm eff}/\hbar \ket{\phi}$ with  
\begin{align}
H_{\rm eff} &= \mathcal{U}^{\dagger}H\mathcal{U} - i\hbar \mathcal{U}^{\dagger} \frac{d\mathcal{U}}{dt} \nonumber\\
& = -\hbar \delta\omega_p \ket{1}\bra{1} - \hbar \left( \delta\omega_p - \delta\omega_c \right)\ket{2}\bra{2} \nonumber\\
& \quad - \hbar \left(\Omega_p \ket{1}\bra{3} + \Omega_c \ket{1}\bra{2} + h.c. \right).
\label{Ham_eff}
\end{align}

When $\delta\omega_p = \delta\omega_c = \delta\omega$ is assumed for simplicity, the energy eigenvalues and eigenstates of $H_{\rm eff}$ are 
\begin{align}
\bar{\lambda}_0 &= 0 \nonumber\\
\bar{\lambda}_\pm &= 0.5\hbar\left(\delta\omega \pm \sqrt{\delta\omega^2 + 4(\Omega_p^2 + \Omega_c^2)} \right), 
\end{align}
and 
\begin{align}
\ket{0} &= \cos \theta\ket{3} - \sin \theta\ket{2},  \nonumber\\
\ket{-} &= \sin \theta\cos \phi\ket{3} + \cos \theta \cos \phi\ket{2} - \sin \phi\ket{1} ,  \nonumber\\
\ket{+} &= \sin \theta\sin \phi\ket{3} + \cos \theta \sin \phi\ket{2} + \cos \phi\ket{1},
\label{eqn:eigenstates}
\end{align}
where the mixing angles $\theta$ and $\phi$ are defined as 
\begin{align}
\theta &= \tan^{-1}\left(\Omega_p/\Omega_c\right)\nonumber\\
\phi &= 0.5\tan^{-1}\left(2\sqrt{\Omega_p^2 + \Omega_c^2}\Big/\delta\omega\right).
\end{align}

Under the two-photon resonance condition ($\delta\omega=0$), the eigenstate $\ket{0}$, a coherent superposition between the states $\ket{2}$ and $\ket{3}$, of the effective Hamiltonian (Eq.~(\ref{Ham_eff})) has zero eigenvalue. 
Hence, the state $\ket{0}$ is a \emph{dark state} that does not evolve with time, and is decoupled from the applied fields. 
Now the spontaneous emission from the state $\ket{1}$ always populates the quantum states $\ket{2}$ and $\ket{3}$. 
Therefore, irrespective of the initial condition, the atomic population is trapped in the dark state $\ket{0}$ for an extended period of time, $t\gg 1/\gamma$. This corresponds to the CPT.
\\

{\bf EIT.} For a strong control field ($\xi=\Omega_c/\Omega_p\gg 1$) and $\delta\omega = 0$, 
a coherent superposition of states $\ket{1}$ and $\ket{2}$, produces the dressed states $\ket{\pm}$, without affecting the state $\ket{3}(=\ket{0})$ (Fig.~\ref{EIT}B). 
The three energy eigen-states and corresponding eigenvalues (inside parenthesis) are obtained as
\begin{align}
\ket{0} &= \ket{3} \qquad (\bar{\lambda}_0 = 0),  \nonumber\\
\ket{\pm} &= \frac{1}{\sqrt{2}}\left( \ket{2} \pm \ket{1} \right) \qquad (\bar{\lambda}_\pm = \pm \hbar \Omega_c).
\end{align}
In this case, the transition amplitude at the resonant probe frequency ($\delta\omega_p= 0 $) between the ground state $\ket{0}=\ket{3}$ to the dressed states $\ket{\pm}$ can be written as $\bra{3}\vec{d}\ket{+} + \bra{3}\vec{d}\ket{-} \simeq \vec{d}_{32} + \vec{d}_{31} + \vec{d}_{32} - \vec{d}_{31} = 2\vec{d}_{32}=0$ because of the electric dipole selection rule that disallows the transition between $\ket{2}$ and $\ket{3}$ ($\vec{d}_{32}=0$). 
Consequently, all the population is effectively confined in the dark state $\ket{0}$. 
At $\delta\omega_p = 0$, the media is transparent to the pulse, and  does not absorb the probe pulse. 
This strong control field-induced ($\Omega_c\gg \Omega_p$) conversion of an absorptive medium to a transparent one is termed the electromagnetically induced transparency (EIT) \cite{scully_zubairy_1997}. 
The EIT creates the destructive interference between the transition pathways $\ket{3} \rightleftharpoons \ket{1}$ and $\ket{2} \rightleftharpoons \ket{1} \rightarrow \ket{3}$. 

The energy gap between the dressed states is  $2\hbar\Omega_c$. Then, the conditions for the perfect resonance between $\ket{0}$ and $\ket{\pm}$ appears when $\delta\omega_p =  \pm\Omega_c$, resulting in the complete absorption of probe pulse, giving rise to the Aulter-Townes absorption doublet \cite{scully_zubairy_1997}. 
The off-resonant probe pulse ($\delta\omega_p\approx 1$) engenders the absorption doublet where again the dispersion becomes zero ($\rho_{13}^R = 0$), but this time the absorption ($\rho_{13}^I$) is maximized. 
\\

\setcounter{figure}{0}
\renewcommand{\thefigure}{B\arabic{figure}}
\setcounter{equation}{0}
\renewcommand{\theequation}{B\arabic{equation}}

\subsection{Evolution equation}
The total Hamiltonian in the presence of an external field is expressed as \cite{scully_zubairy_1997,Carmichael}
\begin{align}
H&=H_S+H_{\rm int} +H_B+H_{SB},
\label{Hamiltonian}
\end{align}
where
\begin{align}
H_S &= \hbar\left( \omega_1\ket{1}\bra{1} + \omega_2\ket{2}\bra{2} + \omega_3\ket{3}\bra{3} \right) \nonumber\\
H_{\rm int} &= - \vec{d}_2\cdot\vec{E}_c - \vec{d}_3\cdot\vec{E}_p\nonumber\\
H_B &= \sum_{\textbf{k},\lambda} \hbar\omega_{\textbf{k},\lambda}b_{\textbf{k},\lambda}^{\dagger}b_{\textbf{k},\lambda} \nonumber\\
H_{SB} &= \sum_{\textbf{k},\lambda}\hbar\left[\left(g_{\textbf{k},\lambda}^{\ast}\right)_{12}b_{\textbf{k},\lambda}^{\dagger}\ket{2}\bra{1} + \left(g_{\textbf{k},\lambda}\right)_{12}b_{\textbf{k},\lambda}\ket{1}\bra{2} \right. \nonumber\\
& \left.+\left(g_{\textbf{k},\lambda}^{\ast}\right)_{13}b_{\textbf{k},\lambda}^{\dagger}\ket{3}\bra{1} + \left(g_{\textbf{k},\lambda}\right)_{13}b_{\textbf{k},\lambda}\ket{1}\bra{3}\right], 
\label{Hamiltonian1}
\end{align}
with $H_S$ denoting the $\Lambda$-system, $H_B$ background quantized radiation, and $H_{SB}$ the interaction between the system and radiation. 
The control and probe fields, $\vec{E}_\alpha(t) = \hat{e}_\alpha\zeta_\alpha (e^{i\omega_\alpha t} +e^{-i\omega_\alpha t})$ with $\alpha=c$ and $p$ where $\hat{e}_\alpha$ is the unit vector representing the direction of polarization and  
$\zeta_\alpha$ denotes the amplitude, interact with the $\Lambda$-system via the interaction energy Hamiltonian $H_{\rm int}=- \vec{d}_2\cdot\vec{E}_c - \vec{d}_3\cdot\vec{E}_p$, 
inducing the excitations of $\ket{2}\rightarrow\ket{1}$ and $\ket{3}\rightarrow\ket{1}$, respectively. 
The transition dipole operator is given by 
$\vec{d} = \vec{d}_2+\vec{d}_3=\left(\vec{d}_{12}\ket{1}\bra{2} + \vec{d}_{21}\ket{2}\bra{1}\right) + \left(\vec{d}_{13}\ket{1}\bra{3} + \vec{d}_{31}\ket{3}\bra{1}\right)$ with the dipole matrix elements, $\vec{d}_{ij}$. 
Since the transition between $\ket{2}$ and $\ket{3}$ 
is effectively forbidden, $\vec{d}_{23} = \vec{d}_{32} \approx 0$.
The summation $\sum_{\textbf{k},\lambda}$ extends over the wavevector $\textbf{k}$ and polarization $\lambda$. 
The symbols, $b_{\textbf{k},\lambda}^{\dagger}$ and $b_{\textbf{k},\lambda}$ denote the creation and annihilation operators of the harmonic oscillators of angular frequency $\omega_k$ constituting the reservoir. The dipole coupling constant, $\left(g_{\textbf{k},\lambda}\right)_{1j}\equiv -i\sqrt{\omega_k/2\hbar\varepsilon_0V}\hat{e}_{\textbf{k},\lambda}\cdot\vec{d}_{1j}$ for $j \in 2,3$, 
contains the information of polarization $\hat{e}_{\textbf{k},\lambda}$, quantization volume $V$ and vacuum permittivity $\varepsilon_0$.

The density matrix for the total system, $\rho_\text{tot}(t)$, evolves with time, obeying the von Neumann equation, $d\rho_\text{tot}(t)/dt=-\frac{i}{\hbar}[H,\rho_\text{tot}]$. 
In the framework of Lindblad approach, the reduced density matrix after tracing out the bath degrees of freedom obeys the following evolution equation. 
\begin{align}
&\frac{d\rho(t)}{dt}= - \frac{i}{\hbar}[H_{S} +H_{\rm int}, \rho]\nonumber\\
&+\gamma_{12}(\bar{n}_{12}+1)\left(\ket{2}\bra{1}\rho\ket{1}\bra{2} - \frac{1}{2}\{\ket{1}\bra{1},\rho\}_+\right)\nonumber\\
&+\gamma_{12}\bar{n}_{12}\left(\ket{1}\bra{2}\rho\ket{2}\bra{1} - \frac{1}{2}\{\ket{2}\bra{2},\rho\}_+\right)\nonumber\\
&+\gamma_{13}(\bar{n}_{13}+1)\left(\ket{3}\bra{1}\rho\ket{1}\bra{3} -\frac{1}{2}\{\ket{1}\bra{1},\rho\}_+\right)\nonumber\\
&+\gamma_{13}\bar{n}_{13}\left(\ket{1}\bra{3}\rho\ket{3}\bra{1} -\frac{1}{2}\{\ket{3}\bra{3},\rho\}_+\right),
\label{master}
\end{align}
where $\gamma_{1j} =4\omega_{1j}^3|d_{1j}|^2/(3\hbar c^3)$ is the spontaneous decay rate from the excited state $\ket{1}$ to the ground state $\ket{j}$ ($j=2,3$), $\bar{n}_{1j} = (e^{\beta\hbar\omega_{1j}} - 1)^{-1}$ is the mean number of thermal photons with $\beta = 1/k_BT$, and  
$\{A,B\}_+\equiv AB+BA$ denotes the anti-commutator. 

After eliminating the terms violating the energy conservation \cite{scully_zubairy_1997}, which amounts to taking the rotating wave approximation (RWA), the energy hamiltonian for the light-matter interaction is simplified to  
\begin{align}
H_{\rm int}
&\simeq -\hbar\Omega_c\left(e^{-i\omega_c t}\ket{1}\bra{2} + e^{i\omega_c t}\ket{2}\bra{1}\right)\nonumber\\
&-\hbar\Omega_p\left(e^{-i\omega_p t}\ket{1}\bra{3} + e^{i\omega_p t}\ket{3}\bra{1} \right)
\label{H_SE}
\end{align}
where $\Omega_c = \zeta_c |\hat{e}_c\cdot \vec{d}_{12}|/\hbar $ and $\Omega_p = \zeta_p |\hat{e}_p\cdot \vec{d}_{13}|/\hbar $ corresponds to the driving (Rabi) frequencies. 
With $H_S$ in Eq.~(\ref{Hamiltonian1}), $H_{\rm int}$ in Eq.~(\ref{H_SE}), and transformations into rotating frame which lead to $\rho_{ii} \rightarrow \tilde{\rho}_{ii}$, $\rho_{12} \rightarrow \tilde{\rho}_{12}e^{-i\omega_c t}$, $\rho_{13} \rightarrow \tilde{\rho}_{13}e^{-i\omega_p t}$, and $\rho_{23} \rightarrow \tilde{\rho}_{23}e^{-i(\omega_p- \omega_c) t}$ (see Appendix C), 
the transformed matrix elements $\tilde{\rho}_{ij}$'s evolve with time as follows. 
\begin{align}
\frac{d\tilde{\rho}_{22}}{d\tau}  &=\gamma(\bar{n}_{12} + 1)\tilde{\rho}_{11} + i\Omega_c\tilde{\rho}_{12} - i\Omega_c\tilde{\rho}_{21} - \gamma\bar{n}_{12}\tilde{\rho}_{22}  \nonumber\\
\frac{d\tilde{\rho}_{33}}{d\tau} &=(\bar{n}_{13} + 1)\tilde{\rho}_{11} + i\Omega_p\tilde{\rho}_{13} - i\Omega_p\tilde{\rho}_{31} - \bar{n}_{13}\tilde{\rho}_{33}  \nonumber\\
\frac{d\tilde{\rho}_{12}}{d\tau} &=
-i\Omega_c\tilde{\rho}_{11} + \left[i\delta\omega_c - \frac{\gamma}{2}(2\bar{n}_{12} + 1) - \frac{(\bar{n}_{13} + 1)}{2} \right]\tilde{\rho}_{12}\nonumber\\
& + i\Omega_c\tilde{\rho}_{22} + i\Omega_p\tilde{\rho}_{32}\nonumber\\
\frac{d\tilde{\rho}_{13}}{d\tau} &=-i\Omega_p\tilde{\rho}_{11} + \left[i\delta\omega_p - \frac{\gamma}{2}(\bar{n}_{12} + 1) - \frac{(2\bar{n}_{13} + 1)}{2} \right]\tilde{\rho}_{13}\nonumber\\
& + i\Omega_c\tilde{\rho}_{23} + i\Omega_p\tilde{\rho}_{33} \nonumber\\
\frac{d\tilde{\rho}_{23}}{d\tau} &= i\Omega_c\tilde{\rho}_{13} - i\Omega_p\tilde{\rho}_{21}\nonumber\\
& + \left[ i(\delta\omega_p - \delta\omega_c) - \frac{(\gamma\bar{n}_{12} + \bar{n}_{13})}{2} \right]\tilde{\rho}_{23},
\label{dynamical_equation2}
\end{align}
where the equations are rescaled with $\gamma_{13}$, redefining the parameters and variables, such that $\tau\equiv\gamma_{13} t$, $\gamma\equiv\gamma_{12}/\gamma_{13}$. 
Hereafter, we implicitly assume that all the rates including $\Omega_c$, $\Omega_p$, $\delta\omega_c$, and $\delta\omega_p$  are those scaled with $\gamma_{13}$, e.g.,  
$\Omega_c/\gamma_{13}\rightarrow \Omega_c$, $(\omega_c - \omega_{12})/\gamma_{13}\rightarrow \delta\omega_c$ and so forth. The equations for the remaining elements are obtained from the constraints $\sum_i\rho_{ii}=1$ and 
$\rho_{ji}=\rho_{ij}^\ast$ for $i\neq j$.


\setcounter{figure}{0}
\renewcommand{\thefigure}{C\arabic{figure}}
\setcounter{equation}{0}
\renewcommand{\theequation}{C\arabic{equation}}

\subsection{Transformation to the rotating frame}
The following operation transforms the state vector $\ket{\phi}$ in the rotating frame into the one in the stationary frame ($\ket{\psi}$). 
\begin{align}
\ket{\psi} = \mathcal{U}(t)\ket{\phi},
\label{rot_trans}
\end{align}
with $\mathcal{U}(t)=e^{-i\omega_p t \ket{1}\bra{1} - i(\omega_p - \omega_c)t \ket{2}\bra{2}}$. 
Then, the density matrix $\tilde{\rho}=\ket{\phi}\bra{\phi}$ in the rotating frame is transformed into the one in the stationary frame via 
$\ket{\psi}\bra{\psi}(=\rho)= \mathcal{U}\ket{\phi}\bra{\phi}\mathcal{U}^{\dagger}(=\mathcal{U}\tilde{\rho}\mathcal{U}^{\dagger})$.   

The Baker-Campbell-Hausdorff formula, 
\begin{align}
e^{s\hat{A}}\hat{B}e^{-s\hat{A}} = \hat{B} + \dfrac{s}{1!}[\hat{A},\hat{B}] + \dfrac{s^2}{2!}[\hat{A},[\hat{A},\hat{B}]] \cdots\nonumber
\end{align}
enables one to rewrite the diagonal elements as $\tilde{\rho}_{jj} = \rho_{jj}$, and the off-diagonal elements as
$\tilde{\rho}_{12} = \rho_{12}e^{i\omega_c t}$, 
$\tilde{\rho}_{13} = \rho_{13}e^{i\omega_p t}$, and  
$\tilde{\rho}_{23} = \rho_{23}e^{i(\omega_p - \omega_c) t}$.

\setcounter{figure}{0}
\renewcommand{\thefigure}{D\arabic{figure}}
\setcounter{equation}{0}
\renewcommand{\theequation}{D\arabic{equation}}

\subsection{The method of cumulant generating function}
{\color{black}In order to calculate the current ($\langle n(\tau)\rangle$) and its fluctuations (${\rm var}[n(\tau)]$), we employ the method of cumulant generating function. 

We start by defining the cumulant generating function $\mathcal{G}(z,\tau)$ as follows:}  
\begin{align}
\mathcal{G}(z,\tau)=\ln{\langle e^{zn}\rangle}=\ln{\sum_nP(n,\tau)e^{zn}},   
\end{align}
which allows one to calculate the $k$-th cumulant
\begin{align}
\langle\langle n^k\rangle\rangle(\tau)=\frac{\partial^k \mathcal{G}(z,\tau)}{\partial z^k}\Big|_{z=0}. 
\label{eqn:cumulant}
\end{align}
{\color{black}Here, $P(n,\tau)\equiv \tilde{\rho}_{11}(n,\tau)+\tilde{\rho}_{22}(n,\tau)+\tilde{\rho}_{33}(n,\tau)$ with a normalization condition $\sum_{n=-\infty}^{\infty}P(n,\tau)=1$ 
denotes the probability that $n$ net photons have been processed by the three states of the $\Lambda$-system and eventually emitted to the environment for the time duration $\tau$. 
The terms, $\tilde{\rho}_{11}(n,\tau)$, $\tilde{\rho}_{22}(n,\tau)$, and $\tilde{\rho}_{33}(n,\tau)$ are the population terms of the reduced density matrix $\tilde{\rho}(n,\tau)$ that satisfies the $n$-resolved master equation, which is explained below (see Eq.~(\ref{eqn:n-resolved})). 
}   

The vectorized form of the reduced density matrix in Fock-Liouville space, 
$\tilde{\varrho}=(\tilde{\rho}_{11},\tilde{\rho}_{12},\tilde{\rho}_{13},\tilde{\rho}_{21},\tilde{\rho}_{22},\tilde{\rho}_{23},\tilde{\rho}_{31},\tilde{\rho}_{32},\tilde{\rho}_{33})^T$ obeys the Liouville equation \begin{align}
\partial_\tau\tilde{\varrho}(\tau)=\mathcal{L}\tilde{\varrho}(\tau), 
\label{eqn:Liouville_Eq}
\end{align}
where $\mathcal{L}$ is the Liouvillian super-operator expressed as $9\times9$ matrix, and 
formally evolves with time as $\tilde{\varrho}(\tau)=e^{\mathcal{L}\tau}\tilde{\varrho}(0)$.  
The vector $\tilde{\varrho}(\tau)$ is decomposed into $\tilde{\varrho}(n,\tau)$, such that  
$\tilde{\varrho}(\tau)=\sum_{n=-\infty}^{\infty}\tilde{\varrho}(n,\tau)$ with $\tilde{\varrho}(n,\tau)$ satisfying the $n$-resolved master equation \cite{flindt2010PRB}
\begin{align}
\partial_\tau\tilde{\varrho}(n,\tau)&=\mathcal{L}_{0}\tilde{\varrho}(n,\tau)\nonumber\\
&+\mathcal{L}_+\tilde{\varrho}(n-1,\tau)+\mathcal{L}_-\tilde{\varrho}(n+1,\tau),	
\label{eqn:n-resolved}
\end{align}
where the generators $\mathcal{L}_{+}$ and $\mathcal{L}_-$ are the off-diagonal element of the $\mathcal{L}$ corresponding to the emissions ($\mathcal{L}_{22,11}$, $\mathcal{L}_{33,11}$) and absorption ($\mathcal{L}_{11,22}$, $\mathcal{L}_{11,33}$), respectively, and  
$\mathcal{L}_{0}$ is for the rest of the elements. 
Discrete Laplace transform $\hat{\varrho}_z(\tau)=\sum_n\tilde{\varrho}(n,\tau)e^{zn}$, which satisfies $\lim_{z\rightarrow0}\hat{\varrho}_z(\tau)=\tilde{\varrho}(\tau)$, casts Eq.~(\ref{eqn:n-resolved}) into \begin{align}
\partial_\tau\hat{\varrho}_z(\tau)=\mathcal{L}(z)\hat{\varrho}_z(\tau)
\end{align}
with {\color{black}the modified super-operator in Laplace space} $\mathcal{L}(z)\equiv \mathcal{L}_{0}+e^z\mathcal{L}_++e^{-z}\mathcal{L}_-$.   
{\color{black}Specifically,} 
\footnotesize
\begin{widetext}
\begin{align}
\mathcal{L}(z) &\equiv \begin{bmatrix} -A_{1} & -i\Omega_c & -i\Omega_p & i\Omega_c & \gamma\bar{n}_{12}e^{-z} & 0 & i\Omega_p & 0 & \bar{n}_{13}e^{-z}\\
-i\Omega_c & i\delta\omega_c - A_{2} & 0 & 0 & i\Omega_c & 0 & 0 & i\Omega_p & 0 \\
-i\Omega_p & 0 & i\delta\omega_p - A_{3} & 0 & 0 & i\Omega_c & 0 & 0 & i\Omega_p \\
i\Omega_c & 0 & 0 & -i\delta\omega_c - A_{2} & -i\Omega_c & -i\Omega_p & 0 & 0 & 0\\
\gamma(\bar{n}_{12}+1)e^{z} & i\Omega_c & 0 & -i\Omega_c & -\gamma\bar{n}_{12} & 0 & 0 & 0 & 0 \\
0 & 0 & i\Omega_c & -i\Omega_p & 0 & i\delta\omega_{pc} - A_{6} & 0 & 0 & 0 \\
i\Omega_p & 0 & 0 & 0 & 0 & 0 & -i\delta\omega_p - A_{3} & -i\Omega_c & -i\Omega_p\\
0 & i\Omega_p& 0 & 0 & 0 & 0 & -i\Omega_c & -i\delta\omega_{pc} - A_{6} & 0\\
(\bar{n}_{13}+1)e^{z} & 0 & i\Omega_p & 0 & 0 & 0 & -i\Omega_p & 0 &  -\bar{n}_{13}  \end{bmatrix},
\end{align}
\end{widetext}
\normalsize
with  
$\delta\omega_{pc} = \delta\omega_{p} - \delta\omega_{c}$, 
$A_{1} = \gamma(\bar{n}_{12} + 1) + (\bar{n}_{13} + 1)$, 
$A_{2} = \gamma(2\bar{n}_{12} + 1)/2 - (\bar{n}_{13} + 1)/2$, 
$A_{3} = \gamma(\bar{n}_{12} + 1)/2 - (2\bar{n}_{13} + 1)/2$, and 
$A_{6} = (\gamma\bar{n}_{12} + \bar{n}_{13})/2$.  
{\color{black}Note that $\mathcal{L}(z)$ at $z=0$ reduces to the original Liouvillian super-operator $\mathcal{L}$ of the Liouville equation (Eq.~(\ref{eqn:Liouville_Eq})), namely, $\mathcal{L}(0)=\mathcal{L}$.} 

{\color{black}The $\hat{\varrho}_z(\tau)$ can be formally solved, and it can be approximated using the largest eigenvalue $\lambda_0(z)$ of the modified super-operator $\mathcal{L}(z)$, which satisfies $\lambda_0(z)>\lambda_1(z)>\cdots>\lambda_{8}(z)$, as follows 
\begin{align}
\hat{\varrho}_z(\tau)&=\sum_{n=-\infty}^{\infty}\tilde{\varrho}(n,\tau)e^{zn}=e^{\mathcal{L}(z)\tau}\hat{\varrho}_z(0)\nonumber\\
&\approx e^{\lambda_0(z)\tau}\tilde{\rho}^{ss}+\cdots,  
\label{eqn:approx_large}
\end{align}
Therefore, it follows from Eq.~(\ref{eqn:approx_large}) that for $\tau\gg 1$, $\ln{\hat{\varrho}_z(\tau)}=\ln{\sum_{n=-\infty}^{\infty}\tilde{\rho}(n,\tau)e^{zn}}\sim\lambda_0(z)\tau$, and hence  
\begin{align}
\mathcal{G}(z,\tau)=\ln{\sum_{n=-\infty}^{\infty}P(n,\tau)e^{zn}}\sim\lambda_0(z)\tau.
\label{eqn:Generating_limt}
\end{align} 
Therefore, Eq.~(\ref{eqn:Generating_limt}) along with Eq.~(\ref{eqn:cumulant}) 
offers the $k$-th cumulant of the current at steady states
\begin{align}
\lim_{\tau\rightarrow\infty}\frac{\langle\langle n^k\rangle\rangle(\tau)}{\tau}=\frac{\partial^k\lambda_0(z)}{\partial z^k}\Big|_{z=0}.  
\label{eqn:differentiation}
\end{align}

In principle, Eq.~(\ref{eqn:differentiation}) can be evaluated by calculating the largest eigenvalue $\lambda_0(z)$ of $\mathcal{L}(z)$ explicitly. 
However, drastic simplification in algebra can be made 
by using the following two properties: (i) Along with $\lambda_k(z)$ ($k=1,2,\ldots,8$), $\lambda_0(z)$ is a root of the characteristic polynomial {\color{black}(or the secular equation)} of $\mathcal{L}(z)$ 
\begin{align}
0&=\det | \lambda(z)\mathcal{I} - \mathcal{L}(z)| = \sum\limits_{n = 0}^9 a_n(z)\lambda^n(z)\nonumber\\
&=a_0(z)+a_1(z)\lambda(z)+\cdots a_9(z)\lambda^9(z);
\label{pol}
\end{align}
(ii) $\lambda_0(0)=0$, albeit $\lambda_{k\neq0}(0)\neq 0$, since $\hat{\rho}_{z}(\tau)\Big|_{z=0}$ should converge to the steady state value at $\tau\gg 1$, i.e., 
$\hat{\rho}_{z}(\infty)\Big|_{z=0}\sim \tilde{\rho}^{ss}$.
Eq.~(\ref{pol}) differentiated with respect to $z$ and evaluated at $z = 0$ yields $a_0^\prime(0) + a_1(0)\lambda_0^\prime(0) = 0$, and $a_0^{\prime\prime}(0) + a_1^{\prime}(0)\lambda_0^{\prime}(0) + a_1(0)\lambda_0^{\prime\prime}(0) + 2a_2(0)(\lambda_0^\prime(0))^2 = 0$. 
Therefore, the average photon current and fluctuations due to radiative transitions can be expressed in terms of the coefficients of the characteristic polynomial, $a_0(z)$, $a_1(z)$, $a_2(z)$ and their derivatives at $z=0$ as follows \cite{bruderer_inverse_2014}}
\begin{widetext}
\begin{align}
J_{\rm ph} &= \lim_{\tau\rightarrow\infty}\frac{\langle n\rangle(\tau)}{\tau}=\lambda_0'(0)=
-\frac{a_0^{\prime}(0)}{a_1(0)} \nonumber \\
D_{\rm ph} &= \lim_{\tau\rightarrow\infty}\frac{\langle\langle n^2\rangle\rangle(\tau)}{\tau}=\lambda_0''(0)=-\frac{\left[ a_0^{\prime\prime}(0) + 2 a_1^{\prime}(0) \lambda_0^{\prime}(0) + 2 a_2(0) (\lambda_0^{\prime}(0))^2\right]}{a_1(0)}\nonumber\\
\mathcal{F}&=\frac{D_{\rm ph}}{J_{\rm ph}}=\frac{a_0^{\prime\prime}(0)}{a_0^{\prime}(0)}\left[ 1 + \frac{2(a_0^{\prime}(0))^2a_2(0) - 2a_0^{\prime}(0)a_1(0)a_1^{\prime}(0)}{a_0^{\prime\prime}(0)(a_1(0))^2} \right]. 
\label{eqn:detailed_exp}
\end{align}
\end{widetext}

\setcounter{figure}{0}
\renewcommand{\thefigure}{E\arabic{figure}}
\setcounter{equation}{0}
\renewcommand{\theequation}{E\arabic{equation}}

\subsection{Populations, coherences, and Fano factor}
The general expressions for the density matrix elements at steady states are too lengthy to display; however, for the case of resonant control pulse ($\delta\omega_c = 0$) 
with $\mathcal{A}\gg 1$ (or $\bar{n} \sim 0$), they are significantly simplified at steady state and written in a manageable form.  
\begin{widetext}
\begin{align}
\tilde{\rho}_{11}&=\frac{4 (\gamma +1) \Omega _c^2 \Omega _p^2\delta \omega _p^2 }{\mathcal{D}} \nonumber\\
\tilde{\rho}_{22}&=\frac{\Omega _p^2 \left[
\gamma\left\{(\gamma+1)^2+4\Omega_c^2\right\}\delta\omega_p^2+4 (\Omega _c^2+\Omega_p^2)(\Omega_c^2+\gamma\Omega_p^2)
\right]}{\mathcal{D}}\nonumber\\
\tilde{\rho}_{33}&=\frac{\Omega _c^2 \left[4 \delta \omega _p^4+ \left\{(\gamma +1)^2-8 \Omega _c^2+4 \Omega _p^2\right\}\delta \omega _p^2+4 \left(\Omega _c^2+\Omega _p^2\right) \left(\Omega _c^2+\gamma  \Omega _p^2\right)\right]}{\mathcal{D}}\nonumber\\
\rho _{12}^R&=-\frac{4 \Omega _c \Omega _p^2 \left(\Omega _c^2+\gamma  \Omega _p^2\right)\delta \omega _p }{\mathcal{D}}\nonumber\\
\rho _{12}^I&=\frac{2 \gamma  (\gamma +1) \Omega _c  \Omega _p^2\delta \omega _p^2}{\mathcal{D}}\nonumber\\
\rho _{13}^R&=\frac{4 \Omega _c^2 \Omega _p \left(\Omega _c^2+\gamma  \Omega _p^2-\delta \omega _p^2\right)\delta \omega _p }{\mathcal{D}}\nonumber\\
\rho _{13}^I&=\frac{2 (\gamma +1) \Omega _c^2  \Omega _p\delta \omega _p^2}{\mathcal{D}}\nonumber\\
\rho _{23}^R&=\frac{4 \Omega _c \Omega _p \left[\Omega _c^2 \delta \omega _p^2-\left(\Omega _c^2+\Omega _p^2\right) \left(\Omega _c^2+\gamma  \Omega _p^2\right)\right]}{\mathcal{D}}\nonumber\\
\rho _{23}^I&=-\frac{2 (\gamma +1) \left(\Omega _c^2+\gamma  \Omega _p^2\right)\Omega _c \Omega _p \delta \omega _p }{\mathcal{D}}
\label{coh_real}
\end{align}
with $\mathcal{D} =4 \Omega _c^2 \delta \omega _p^4+ \left[\gamma  (\gamma +1)^2 \Omega _p^2+(\gamma +1) \left(\gamma +1+8 \Omega _p^2\right) \Omega _c^2-8 \Omega _c^4\right] \delta \omega _p^2+4 \left(\Omega _c^2+\Omega _p^2\right)^2 \left(\Omega _c^2+\gamma  \Omega _p^2\right)$. 

The coefficients of the characteristic polynomial of $\mathcal{L}(z)$ (Eq.~(\ref{pol})) at $z=0$, which are required for evaluating the quantities in Eq.~(\ref{eqn:detailed_exp}), are obtained as follows. 
\begin{align}
a_0^\prime(0)&=a_0^{\prime\prime}(0)=(\gamma +1)^3 \Omega _c^2  \Omega _p^2\delta \omega _p^2, \nonumber\\
a_1(0)&=- (\gamma +1) \left[ \Omega _c^2 \delta \omega _p^4+ \left\{(\gamma +1) \Omega _c^2 \left(\gamma +8 \Omega _p^2+1\right)-8 \Omega _c^4+\gamma  (\gamma +1)^2 \Omega _p^2\right\}(\delta \omega _p^2/4)+ \left(\Omega _c^2+\Omega _p^2\right)^2 \left(\Omega _c^2+\gamma  \Omega _p^2\right)\right],\nonumber\\
a_1^\prime(0)&=(\gamma +1) \left[\gamma  \Omega _c^2 \delta \omega _p^4+ \left\{\gamma  \Omega _c^2 \left((\gamma +1)^2-8 \Omega _c^2\right)+(\gamma +1) \Omega _p^2 \left(20 \Omega _c^2+\gamma +1\right)\right\}(\delta \omega _p^2/4)+\left(\Omega _c^2+\Omega _p^2\right)^2 \left(\gamma  \Omega _c^2+\Omega _p^2\right)\right],\nonumber\\
a_2(0)&=\frac{1}{16} \left[-4 \left\{8 (\gamma +2) \Omega _c^2+(\gamma +1)^3\right\}\delta \omega _p^4  \right.\nonumber\\ 
&\left.+\left\{64 (\gamma +2) \Omega _c^4-8 \Omega _c^2 \left(3 (\gamma +1)^2+4 (6 \gamma +7) \Omega _p^2\right)-(\gamma +1) \left((\gamma +1)^4+8 (4 \gamma +1) (\gamma +1) \Omega _p^2+16 \Omega _p^4\right)\right\}\delta \omega _p^2 \right.\nonumber\\ 
&\left.-4 \left(\Omega _c^2+\Omega _p^2\right) \left\{8 (\gamma +2) \Omega _c^4+(\gamma +1) \Omega _c^2 \left((\gamma +1) (\gamma +5)+24 \Omega _p^2\right)+8 (2 \gamma +1) \Omega _p^4+(\gamma +1)^2 (5 \gamma +1) \Omega _p^2\right\}\right].
\label{eqn:a_coeff}
\end{align}

It can be shown that 
\begin{align}
\frac{2(a_0^{\prime}(0))^2a_2(0) - 2a_0^{\prime}(0)a_1(0)a_1^{\prime}(0)}{a_0^{\prime\prime}(0)(a_1(0))^2}= 2\sum_{i<j}(\tilde{\rho}_{ij}^R)^2 - 6 \sum_{i<j}(\tilde{\rho}_{ij}^I)^2+ q(\Omega_c,\Omega_p,\delta\omega_p,\gamma)
 \end{align}
where 
\begin{align}
q(\Omega_c,\Omega_p,\delta\omega_p,\gamma)
= \frac{2q_n}{q_d}
\label{eqn:q_org}
\end{align}
with 
\begin{align}
q_n &= 16 \gamma  \Omega _c^4 \delta \omega _p^8 -8 \gamma  \Omega _c^2 \left[8 \Omega _c^4-\left\{(\gamma +1)^2+2 \Omega _p^2\right\}\Omega _c^2 +(\gamma +1)^2 \Omega _p^2\right]\delta \omega _p^6 \nonumber\\
& + \left[96 \gamma  \Omega _c^8-16 \gamma  \Omega _c^6 \left((\gamma +1)^2-(\gamma +2) \Omega _p^2\right)+(\gamma +1) \Omega _c^4 \left(\gamma  (\gamma +1)^3+4 \gamma  (\gamma +1) \Omega _p^2-32 \Omega _p^4\right) \right.\nonumber\\ 
&\left. -2 \gamma  \Omega _c^2 \Omega _p^2 \left((\gamma +1)^4+6 (\gamma +1)^2 \Omega _p^2+16 \Omega _p^4\right)+\gamma  (\gamma +1)^4 \Omega _p^4\right]\delta \omega _p^4  \nonumber\\
& - 4 \left[16 \gamma  \Omega _c^{10}-2 \gamma  \Omega _c^8 \left\{(\gamma +1)^2-2 (2 \gamma +7) \Omega _p^2\right\}+\Omega _c^6 \Omega _p^2 \left\{\gamma  \left(-\gamma ^3+3 \gamma +2\right)+4 \left(3 \gamma ^2+\gamma +1\right) \Omega _p^2\right\} \right.\nonumber\\ 
&\left. +2 \Omega _c^4 \Omega _p^4 \left\{\left(\gamma ^2+\gamma +1\right) (\gamma +1)^2+2 ((\gamma -3) \gamma +1) \Omega _p^2\right\}+\Omega _c^2 \Omega _p^6 \left(\gamma  \left(\gamma  (2 \gamma +3)-4 \Omega _p^2\right)-1\right)-2 \gamma  (\gamma +1)^2 \Omega _p^8\right]\delta \omega _p^2  \nonumber\\
&+16 \left(\Omega _c^2+\Omega _p^2\right)^2 \left(\Omega _c^2+\gamma  \Omega _p^2\right) \left(\gamma  \Omega _c^6+2 \gamma  \Omega _c^4 \Omega _p^2+2 \Omega _c^2 \Omega _p^4+\Omega _p^6\right) \nonumber\\
q_{d} &= \left[4 \Omega _c^2 \delta \omega _p^4-\left\{8 \Omega _c^4-(\gamma +1)(\gamma +1+8 \Omega _p^2)\Omega _c^2 -\gamma  (\gamma +1)^2 \Omega _p^2\right\}\delta \omega _p^2 +4 \left(\Omega _c^2+\Omega _p^2\right)^2 \left(\Omega _c^2+\gamma  \Omega _p^2\right)\right]^2\nonumber
\end{align}

For $\delta\omega_p=0$, 
\begin{align}
q(\cdot) \mid_{\delta\omega_p=0}&=\frac{32 \left(\Omega _c^2+\Omega _p^2\right)^2 \left(\Omega _c^2+\gamma  \Omega _p^2\right) \left(\gamma  \Omega _c^6+2 \gamma  \Omega _c^4 \Omega _p^2+2 \Omega _c^2 \Omega _p^4+\Omega _p^6\right)}{\left[4 \left(\Omega _c^2+\Omega _p^2\right)^2 \left(\Omega _c^2+\gamma  \Omega _p^2\right)\right]^2}\nonumber\\
&=\frac{2(\gamma\xi^6+2\gamma\xi^4+2\xi^2+1)}{(\xi^2+1)(\xi^2+\gamma)^2}
\label{eqn:q_0}
\end{align}
\end{widetext}
Whereas $q=0$ in a coherently driven TLS \cite{singh2021PRE}, 
$q(\Omega_c,\Omega_p,\delta\omega_p,\gamma)\neq 0$ in the $\Lambda$-system 
contributes to the Fano factor of the transition current.

Although the expressions for $a_0(z)$ and $a_1(z)$ are lengthy and complicated, the total average photon current $J_{\rm ph}$ is straightforwardly decomposed into the two parts, 
$J_{\rm ph}= J_{{\rm ph},12} + J_{{\rm ph},13}$ with 
\begin{align}
J_{{\rm ph},12}&=\gamma(\bar{n}_{12}+1)\tilde{\rho}^{ss}_{11}-\gamma\bar{n}_{12}\tilde{\rho}_{22}^{ss}=2\Omega_c\tilde{\rho}_{12}^{I} 
\label{eqn:j12}
\end{align}
and
\begin{align}
J_{{\rm ph},13}&=(\bar{n}_{13}+1)\tilde{\rho}^{ss}_{11}-\bar{n}_{13}\tilde{\rho}_{33}^{ss}=2\Omega_p\tilde{\rho}_{13}^{I}.
\label{eqn:j13}
\end{align}
The first equalities of Eqs.\ref{eqn:j12} and \ref{eqn:j13} are consistent with the definition of reaction current between two discrete states in classical Markov jump system, and this can also be related with the imaginary part of coherence between the two quantum states, which is called \emph{current-coherence} relation \cite{wu_efficient_2012}). 
Note that at two-photon resonance ($\delta\omega_p=\delta\omega_c=0$) that engenders the \emph{dark state}, the mean current as well as its variance along the two channels vanishes, i.e., $J_{\rm ph}=0$ and $D_{\rm ph}=0$ due to $\rho_{12}^I=\rho_{13}^I=0$ (Eq.~\ref{coh_real}) or $a_0'(0)=a_0''(0)=0$ and $\lambda'(0)=0$ (Eq.~\ref{eqn:a_coeff})); yet the their ratio, the Fano factor of the photon current, $\mathcal{F}=D_{\rm ph}/J_{\rm ph}$, remains finite with its maximal bound, $\mathcal{F}_{\rm max}=3$.

\setcounter{figure}{0}
\renewcommand{\thefigure}{F\arabic{figure}}
\setcounter{equation}{0}
\renewcommand{\theequation}{F\arabic{equation}}

\subsection{Coherent control of dispersion of media}
The probe pulse-induced polarization of the $\Lambda$-system is
quantified with the dipole moment between $\ket{1}$ and $\ket{3}$ per unit volume 
as $\vec{P}_{13} =N\langle \vec{d}_{3} \rangle= \chi_{13}\vec{E}_p$, where $N$ is the number density of atoms. 
$\vec{P}_{13} = \hat{e}_p\zeta_p\chi_{13} e^{-i\omega_p t} + c.c.$, where $\chi_{13}$ is the linear susceptibility of the medium \cite{scully_zubairy_1997}. 
Since $\langle \vec{d}_{3} \rangle = {\rm Tr}(\tilde{\rho}\vec{d})=\tilde{\rho}_{13}\vec{d}_{31} + \tilde{\rho}_{31}\vec{d}_{13}=\rho_{13}e^{i\omega_pt}\vec{d}_{31}+\rho_{31}e^{-i\omega_pt}\vec{d}_{13}\simeq e^{i\omega_pt}\rho_{13}\vec{d}_{31}=\tilde{\rho}_{13}\vec{d}_{31}$, the linear susceptibility can be expressed as $\chi_{13}=|\vec{P}_{13}|/|\vec{E}_p|= N_d\tilde{\rho}_{13}$ with $N_d \equiv N|\vec{d}_{31}|/\zeta_p$. 
For the medium with $|\chi_{13}|\ll 1$, 
the refractive index, dielectric constant and linear susceptibility for the probe field are related with one another in Gaussian units as 
\begin{align}
\eta_{13} (= \sqrt{\epsilon_{13}}) &= \sqrt{1 + 4\pi\chi_{13}} \nonumber\\
&\simeq 1 + 2\pi\chi^{R}_{13} + i2\pi\chi^{I}_{13}.
\end{align}
where $\chi^R$ and $\chi^I$ are the real and imaginary parts of the susceptibility. 
When the probe field, $\vec{E}_p\sim e^{ik_pz}\sim e^{i\beta z}e^{-\alpha z/2}$, passes across the dielectric medium with a wave vector $k_p$, 
\begin{align}
k_p &= \frac{\omega_p}{c}\eta_{13} 
= \underbrace{\frac{\omega_p}{c} \left( 1 + 2\pi\chi^{R}_{13} \right)}_{=\beta} + \frac{i}{2}\underbrace{\frac{\omega_p}{c} 4\pi\chi^{I}_{13}}_{=\alpha}, 
\end{align}
it moves through the medium with a phase velocity $c/(1 + 2\pi\chi^{R}_{13})$, and  
is also attenuated by the medium with an absorption coefficient $\alpha$.
Since $\chi_{13}=N_d\tilde{\rho}_{13}$, the real and imaginary parts of the susceptibility is linked to the dispersion and absorption profiles of the medium, respectively, as $\chi^R_{13} = N_d\rho_{13}^R$ and $\chi^{I}_{13} = N_d \rho_{13}^I$.

\setcounter{figure}{0}
\renewcommand{\thefigure}{G\arabic{figure}}
\setcounter{equation}{0}
\renewcommand{\theequation}{G\arabic{equation}}

\subsection{Relation between $v_g$ and $\mathcal{F}$}
For the case of resonant control pulse ($\delta\omega_c = 0$) with $\mathcal{A}\gg 1$ (or $\bar{n} \sim 0$), 
when $(\partial\rho_{13}^R/\partial\omega_p)_{\delta\omega_p=0}=\Omega_p^{-1}\xi^2/(\xi^2+1)^2$ is inserted to Eq.~(\ref{eqn:vg}), we get an expression of the group velocity in terms of $\xi$. 
\begin{align}
v_g=\frac{c}{1+\dfrac{\mathcal{N} \xi^2}{(\xi^2+1)^2}}
\end{align}
with $\mathcal{N}\equiv 2\pi N_d\omega_p/\Omega_p$. 

For the two-photon resonance ($\delta\omega_c=\delta\omega_p=0$), the Fano factor is contributed only by the real part of coherence between $\ket{2}$ and $\ket{3}$ ($\rho_{23}^R\neq0$) 
while others vanish ($\rho_{12}^R = \rho_{12}^I = \rho_{13}^R = \rho_{13}^R = \rho_{23}^I = 0$), which simplifies $\mathcal{F}$ into  
\begin{align}
\mathcal{F}=1+2\left(\rho_{23}^R\right)^2\Big|_{\delta\omega_p=0}+q(\xi,\gamma)
\label{eqn:Fano_SI}
\end{align}
with
\begin{align}
\left(\rho_{23}^R\right)^2\Big|_{\delta\omega_p=0} &= \frac{\xi^2}{(\xi^2+1)^2} \nonumber\\
q(\xi,\gamma)\Big|_{\delta\omega_p=0}&=\frac{2(\xi^6\gamma + 2 \xi^4\gamma + 2\xi^2+1)}{(\xi^2+1)^2 (\xi^2 + \gamma)}.
\label{eqn:q} 
\end{align}
Insertion of Eq.~(\ref{eqn:q}) into Eq.~(\ref{eqn:Fano_SI}) yields Eq.~(\ref{eqn:Fano}).  

\setcounter{figure}{0}
\renewcommand{\thefigure}{H\arabic{figure}}
\setcounter{equation}{0}
\renewcommand{\theequation}{H\arabic{equation}}

\subsection{Laser power and Rabi frequency}
For a plane wave the average intensity can be expressed as
\begin{align}
\langle I_\alpha \rangle = \frac{c}{8\pi}\zeta_\alpha^2 \qquad \alpha \in c, p.
\end{align}
Now by considering the polarization of incident light parallel to the dipole, we can write $\zeta_\alpha = \hbar\Omega_\alpha/|d_{ij}|$ which yields
\begin{align}
\langle I_\alpha \rangle= \frac{c\hbar^2\Omega_\alpha^2}{8\pi |d_{ij}|^2},
\end{align}
and from the spontaneous decay we know $\left(\hbar/|d_{ij}|\right)^2 = 16\pi^2 h/ 3\gamma_{ij} \lambda_\alpha^3$. 
Thus, we obtain 
the relationship between the average intensity of the laser pulse ($\langle I_\alpha\rangle$), reported in the literature~\cite{hau1999Nature}, 
and other quantities, 
\begin{align}
\langle I_\alpha \rangle&= \frac{ 2 \pi h c \Omega_\alpha^2}{3\gamma_{ij} \lambda_\alpha^3}.
\end{align}

%

\end{document}